# Versatile manipulation of light- and dark-seeking particles on demand

**Zheng Yuan,** [1,2] **Chenchen Zhang,** [1,2] **Yuan Gao,** [1,2] **Wenxiang Yan** [1,2], **Zhi-Cheng Ren,** [1,2] **Xi-Lin Wang,** [1,2] **Jianping Ding,** [1,2,3,*] **and Hui-Tian Wang** [1,2]

[1]*National Laboratory of Solid State Microstructures and School of Physics, Nanjing University, Nanjing 210093, China*
[2]*Collaborative Innovation Center of Advanced Microstructures, Nanjing 210093, China*
[3] *Collaborative Innovation Center of Solid-State Lighting and Energy-Saving Electronics, Nanjing 210093, China.*
*\*Corresponding author: jpding@nju.edu.cn*

**Abstract:** We propose a novel approach to enable the agile manipulation of light- and dark-seeking particles. Our approach involves introducing a two-curvilinear perfect optical vortex beam (TC-POVB) generated by superimposing a pair of curved beams. The TC-POVB exhibits the property of a perfect optical vortex, which means that its size remains constant regardless of its topological charge. Additionally, each curve of the TC-POVB can support a distinct orbital flow density (OFD). This enables the application of torques to produce a dark channel that satisfies the requirements for particle size and drives the revolution or rotation motion of the confined dark-seeking particles. To demonstrate the effectiveness of our approach, we manipulate light- and dark-seeking particles experimentally, making them perform various curvilinear trajectories simultaneously, including moving, revolving, and rotating.

## 1. Introduction

Since its introduction by Ashkin et al. in 1986 [1], optical trapping has found widespread applications in fields such as biological and medical sciences [2,3], material engineering [4], and colloidal interactions [5,6]. Typically, conventional optical tweezers are designed to manipulate high-refractive-index (HRI) microscopic particles, which are drawn toward regions of high light intensity, such as the focal point of strongly focused optical tweezers [7]. Conventional methods, however, exhibit limitations in capturing and manipulating low refractive index (LRI) or light-absorbing particles. These particles are typically repelled by intense light, consequently being confined to the dark region of the beam.

Several methods based on computer-generated holograms (CGHs) have been proposed for manipulating dark-seeking particles [8-13]. For instance, Gahagan et al. achieved successful trapping of dark-seeking particles using a Gaussian vortex beam with a central dark core [8-10]. The manipulation of dark regions between adjacent bright rings of a high-order Bessel beam was later demonstrated [11]. However, as particle diameter increases, trapped particles can escape due to the fixed size of dark regions between adjacent bright rings. Liang et al. proposed an improved method of trapping and rotating large LRI microparticles using a quasi-perfect optical vortex generated by Fourier transforming a high-order quasi-Bessel beam [12]. The dynamics of particles in optical vortices of various topological charges with an unchanged annular radius were investigated due to the perfect optical vortex characteristic of a central dark hollow with a radius that does not depend on the topological charge(TC) and the highest gradient of the field on its boundary [13,14]. However, high-order Bessel beams only allow for circulation motion of particles and do not offer flexible regulation and arrangement of the particle motion path. Recently, Grier et al. demonstrated that the difference-of-Gaussians trap can create a dark central core and trap dark-seeking particles rigidly [15]. However, the requirement for unique CGHs realized on a spatial light modulator (SLM) for each particle movement increases the complexity and inconvenience of the method.

In this study, we introduce a novel class of two-curvilinear perfect optical vortex beams (TC-POVBs), formed by superimposing two curved laser beams with prescribed orbital flow density (OFD) variations along arbitrary trajectories. The TC-POVB structure allows for controllable sizes of the dark regions between the two bright curved beams and the tunable OFD carried by each beam. We experimentally demonstrate the ability of the TC-POVB to manipulate both light- and dark-seeking particles simultaneously. The generated beam can drive particle motion revolution around the dark regions nestled between the bright curved beams and can even trigger rotation owing to the carried OFD of the TC-POVB.

## 2. Principle

### 2.1 Generation of curvilinear focal beams

Let us first consider a focusing process under the paraxial condition. We want to generate a focal beam that conforms its intensity and phase distribution to a predefined two-dimensional curve path represented by $s(t) = [x_0(t), y_0(t)]$ with $t \in [0, T]$. Here the parameter $T$ represents the largest possible value of the variable $t$ that determines the terminals of the curve. The desired focal beam is generated by computing the complex amplitude at the incident plane of a focusing lens, expressed as [16, 17]:

$$\vec{E}(x, y) = \int_0^T g(t) \exp\left\{-\frac{i}{w_0^2}\left[xy_0(t) + yx_0(t)\right]\right\} dt \cdot \hat{e}_p, \quad (1)$$

where $w_0$ is a constant and $\hat{e}_p$ denotes the unit vector of the incident light polarization. The term $g(t)$ in Eq. (1) determines the amplitude and phase to be applied along the prescribed curve, given by:

$$g(t) = a(t) \exp\left[i2\pi m \left[\frac{S(t)}{S(T)}\right]^\alpha\right], \quad (2)$$

Here $S(t) = \int_0^t dl$ and $S(T)$ represents the curve length and serves as a normalization factor. For all cases discussed in this study, $T = 2\pi$. It is important to note that the intensity distribution along the curve is determined by the factor $a(t)$; one special case is when $a(t) = \sqrt{[x_0'(t)]^2 + [y_0'(t)]^2}$, resulting in an uniformly distributed intensity along the trajectory. Furthermore, the parameter $m$ defines the phase accumulation along the entire curve, providing global control of the phase along the curves. In the case of closed curves, the parameter $m$ can be understood as the vortex topological charge, i.e., $m = \oint_\ell \nabla \psi d\ell / 2\pi$ [19]. The parameter $\alpha$ governs the local phase gradient along the curve variation rate and plays a critical role in defining to the local motion of the trapped particle along the curve as will be demonstrated in the particle trapping experiments.

In addition, curved circuits can be elegantly represented using the Superformula expression. This expression, given by $\ell(t) = R_0 \left[\left|\frac{1}{a}cos\left(\frac{n_0}{4}t\right)\right|^{n_2} + \left|\frac{1}{b}sin\left(\frac{n_0}{4}t\right)\right|^{n_3}\right]^{-1/n_1}$, employs a constant parameter $R_0$ across all closed curves discussed and a set of real numbers $q = (a, b, n_0, n_1, n_2, n_3)$ capable of generating a variety of closed polygons having different degrees of symmetry [23].

*2.2 Generation of TC-POVB via tight focusing*

Normally, stable optical trapping of microscopic particles is realized in a tightly focused field, which deviates from the paraxial condition. Therefore, the preceding design for the curvilinear focal beams needs adjustment to accommodate this tight focusing scenario. We propose to construct TC-POVBs within the focal volume, leveraging a technique we previously developed for creating structured vector beams [18]. This is accomplished by computing a CGH using Eq. (1) and integrating it with a vector optical field generator [18*]. Together, The CGH and the vector optical field generator produce an incident field for the focusing lens. This incident field is composite of two vector fields, denoted by $\vec{E}_i = \vec{E}_1 + \vec{E}_2$, with each field having its unique complex amplitude and polarization state.

When the light field $\vec{E}_i$ is incident on the lens, it results in a transmitted field $\vec{E}_t$ at the exit pupil, which can be expressed in the cylindrical coordinate system $(\rho, \phi, z)$ as follows [20]:

$$\vec{E}_t(\theta, \phi) = \sqrt{\cos\theta}\, \mathbf{M}(\theta, \phi)\, \vec{E}_i(x, y), \tag{3}$$

Here $\sqrt{\cos\theta}$ is responsible for ensuring energy conservation throughout the transmission process. The polarization transformation matrix $\mathbf{M}(\theta, \phi)$ in Eq. (4) encapsulates three successive geometric rotation operations on the incident field vector as the refracted ray is bent towards the lens's focus, and can be expressed as

$$\mathbf{M}(\theta, \phi) = \begin{bmatrix} \cos^2\phi\cos\theta + \sin^2\phi & (\cos\theta - 1)\sin\phi\cos\phi & -\sin\theta\cos\phi \\ (\cos\theta - 1)\sin\phi\cos\phi & \cos^2\phi + \sin^2\phi\cos\theta & -\sin\theta\sin\phi \\ \sin\theta\cos\phi & \sin\theta\sin\phi & \cos\theta \end{bmatrix}, \tag{4}$$

The light field near the focal point of the lens can be computed via the Richards-Wolf vectorial diffraction integral, and can furthermore be simplified in terms of Fourier transform as [21, 22]:

$$\vec{E}_f(x, y, z) = \mathcal{F}\{P(\rho)\vec{E}_t(\theta, \phi)\exp(ik_z z)/\cos\theta\}, \tag{5}$$

wherein a multiplicative constant factor has been dropped for clarity. Phase accumulation during propagation along the z-axis is accounted for by the phase factor $\exp(ik_z z)$, and the two-dimensional Fourier transform is represented by $\mathcal{F}\{\cdot\}$. In Eq. (5), $k_z$ is the z component of wave vector, $\mathcal{F}\{\cdot\}$ denotes the 2D Fourier transform, and $P(\rho) = 1$ inside the aperture of lens, and $P(\rho) = 0$ outside the aperture of lens.

*2.3 Flow density of TC-POVB*

We now examine the photonic momentum density distribution within the focal field. This energy flow will governs the dynamical behavior of particles in the focal field. The time-averaged momentum density can be calculated by utilizing the Poynting energy flow vector and can be expressed as [24, 25]

$$\vec{p} = \operatorname{Im}\left[\vec{E}_f^{\,*} \times \left(\nabla \times \vec{E}_f\right)\right], \tag{6}$$

In accordance with Eq. (6), the z component of the angular momentum density of light is given by

$$\vec{J}_z = \vec{r} \times \vec{p}_\perp, \tag{7}$$

where the symbol $\vec{p}_\perp$ represents the transverse component of momentum density $\vec{p}$.

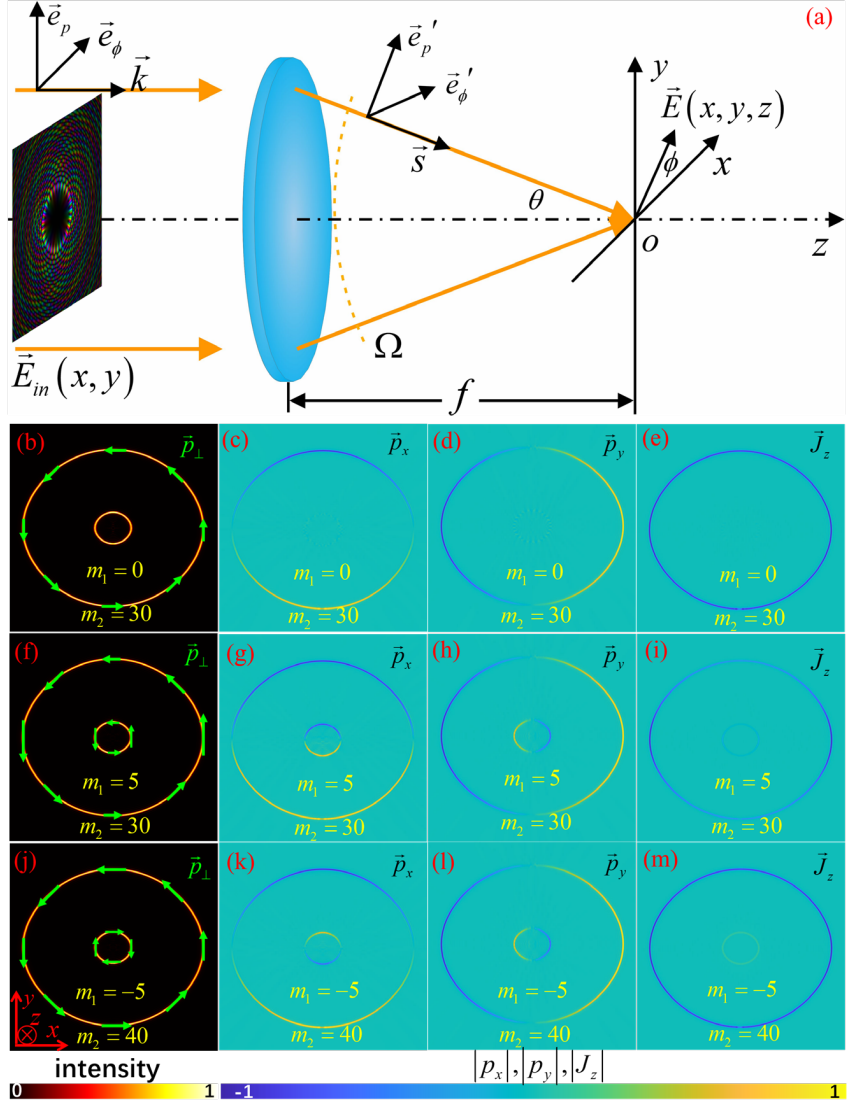

Fig.1 (a) schematic of the focusing geometry. (b)-(m) Simulation of the total intensity distribution of the tightly focused focal field, accounting for momentum density in the transverse plane and the z-component angular momentum density.

The design paths of elliptical TC-POVB curves for $q=(1,1,2,10,6,6)$ are presented in Fig.1, where the radius parameter $R_0$ of the inner ellipse is configured as 0.1, and the outer ellipse's radius parameter is $R_0=0.5$. The symbols $m_1$ and $m_2$ represent the phase TC of the inner and outer curved beams, respectively. According to Eqs (5-7), Figure 1(b-m) displays the total intensity distribution, the transverse component of the momentum density, and the z-component of the angular momentum density. **For more in-depth discussion of momentum density, please refer to Supplementary Material, Part I.** Figs.1(b) depicts the total intensity of TC-POVB in the focal field, while the green vector arrows denote the schematic diagrams of the momentum density distribution in the transverse plane synthesized from Fig. 1 (c) and Fig. 1 (d) as the basis vectors in the x and y directions, respectively. Figs. 1(e) illustrate the distributions of the z components of the angular momentum density. In the first row of Fig. 1, the parameters of TC-POVB are set to $\alpha_1=\alpha_2=1$, $m_1=0$, and $m_2=30$, resulting in an momentum density distribution in the transverse plane equal in magnitude everywhere along the prescribed outer curve trajectory, but its direction is tangent to the curve, imparting a driving force with a counterclockwise rotation near the outer curve due to the momentum density to the captured particles. However, on the small curve trajectory of the inner circle, the momentum density's magnitude is zero, and the nearby particles do not experience the momentum. Figs. 1(f-i) display a scenario where the uniform distribution of momentum density size is disrupted, and the inner curve also bears the momentum. In this case, $\alpha_1=\alpha_2=8$, $m_1 = 5$, and $m_2 = 30$ are set, with the momentum density size of the inner and outer circle increasing gradually from the minimum value in the horizontal direction to the maximum value in the vertical direction cyclically. The dark-seeking particles are driven by the momentum on the inner and outer curve along the potential well simultaneously. The distribution of momentum density on the inner and outer curves is arbitrary, which significantly increases the degrees of freedom in particle motion. In contrast to the preceding designs, Fig.1(j-m) illustrates the inner and outer loops with different signs of $m_1$ and $m_2$, where $m_1=-5$, $m_2=40$, and $\alpha_1=\alpha_2=1$. Fig. 1(j) portrays the momentum density distribution on the transverse plane, synthesized from Fig. 1(j) and Fig. 1(k). The inner ellipse provides a uniform driving force for the dark-seeking particles along the tangent of the curve in the counterclockwise direction, while the outer ellipse provides a uniform driving force along the tangent of the curve in the clockwise direction. It is important to note that Fig. 1(b), Fig. 1(f), and Fig. 1(j) are designed from three structures with different m values, but the size of the total intensity distribution of the focusing field, including the inner and outer rings, remains constant. This momentum density design with perfect characteristics provides a paradigm for the simultaneous manipulation of light-seeking and dark-seeking particles.

## 3. Experiment

### 3.1 Manipulation of dark-seeking nanoparticles

Our experimental setup, illustrated in Figure 3, consists of an inverted microscope. To encode the complex field amplitude described by Equation (1), we use a phase-only holographic grating and a programmable reflective spatial light modulator (SLM) (Holoeye Leto, pixel pitch of 6.4 μm, pixel number of 1920×1080). The collimated green laser beam (532 nm wavelength) was expanded by a factor of 5 before being directed onto the computer-generated hologram. The reflective light that leaves the SLM carries the required complex amplitudes $\vec{E}_{in}(x,y)$ and is directed horizontally at an angle $\theta$ using a blazing grating. This +1st diffraction order term is then allowed to pass through a spatial filter at the focal plane of the first lens, filtering out unwanted 0th order items. A notch filter placed at the focal plane of the 4f system's second lens redirects the trapping beam into the objective lens and prevents saturation of the camera with back-scattered laser light during the subsequent photo and video recording process. The complex fields generated are incident upon the objective lens (Nikon plan, 1.1 NA, 100×), which generates the expected TC-POVB with the prescribed intensity and phase distribution at the focal region. The lens focuses the beam, creating a trap in a sample consisting of both dark-seeking (silver-coated hollow glass spheres with diameters ranging from 2-20 μm) and light-seeking (polystyrene spheres with a diameter of 1.3 μm) nanoparticles dispersed in deionized water. The gas-filled hollow glass sphere can serve as a conventional sample as microparticles due to its thin glass shell relative to its radius [8-12]. However, silver-plated hollow glass spheres with absorbent shells are more suitable as ideal dark-seeking particles because the silver coating repels away bright light [26]. The sample is enclosed in a chamber made by attaching a glass coverslip (0.17 mm thickness) and a glass slide (1.2 mm thickness), and a spacer made of silicone (Henkel, Loctite SI 587) is used to ensure that the samples could be reused for as long as possible.

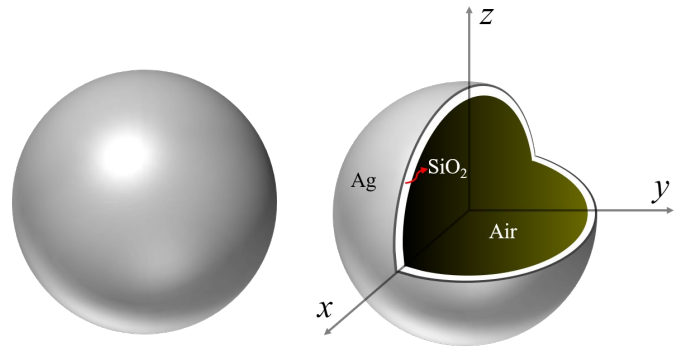

Fig. 2 Schematic representation of a dark-seeking particle

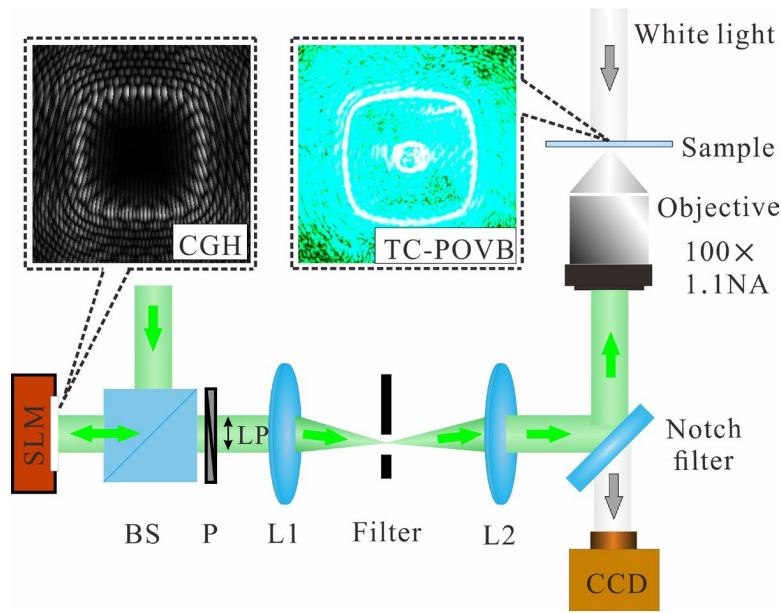

Fig. 3 ketch of the inverted optical trapping microscope. CGH,computer-generated hologram ;SLM, spatial light modulator;BS, beam splitter; P, polarizer; LP, linear polarization; Lens: L1, L2.

Figure. 4(a) displays the focal intensity distribution of the TC-POVB with double ring trajectory by setting $q=(1,1,4,-0.5,2,2)$ and schematically illustrates the transverse forces that a dark-seeking nanoparticle may experience in the focal plane, where the inner and outer diameters are 2.7 um and 13.6 um, respectively. $F_\phi$ denotes the tangent force originating from the OFD that drives the nanoparticle rotation, while $F_\rho$ represents the repulsive force arising from the bright ring to the dark-seeking nanoparticle, directed toward the dark area. In this scenario, the particles are confined in the dark region due to the repulsive forces, and the tangent forces drive the particles to move or spin along a specific trajectory. Notably, the two tangential forces $F_{\rho_1}$ and $F_{\rho_2}$ provided by the inner and outer rings can be clockwise, counterclockwise, or in different directions. Figs. 4(b) demonstrates the effect of transverse OFD carried by each of the inner and outer rings on the motion of the trapped particles in the experiment, and both the inner and outer ring are effective in controlling particle motion velocity. Four inner-loop transverse OFD structures are designed with inner loops m1 = 5, 1, 0, and -1, respectively. In situations featuring high values of TCs associated with small inner curves, particularly when $|m1| \geqslant 5$ in our experimental configuration, the pixelated structure of the spatial light modulator (SLM) results in a deterioration of the quality of the generated light field. This degradation, in turn, has the potential to significantly impact the efficacy of optical trapping.

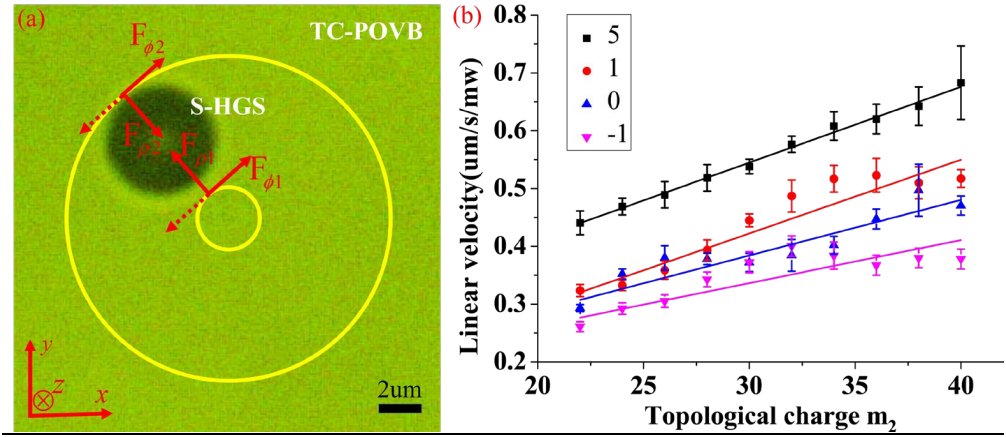

Fig. 4 Silver coated hollow glass spheres with a diameter of 5um are manipulated by TC-POVB(ring trajectory) with respect to the topological charge. (a) Schematic analysis of rotation of a silver coated hollow glass sphere. (b) Normalized linear velocity for power. $F_\phi$: tangent force, $F_\rho$: radial force.

The structures are used to investigate the change of motion of captured dark-seeking particles as the outer-loop transverse OFD is gradually increased from m2 = 0 to 40. To maintain consistency in the experimental environment and measure particle motion efficiently and rigorously, multiple repetitions of measurements were used in each set of experiments. The effective linear velocity was calculated by using focal field power as a normalization factor and measuring the variation of focal field power due to slight changes in the hologram while keeping outgoing laser power constant. The supplementary material provides the focal field power and linear velocity of the particles corresponding to each CGH. At the same outer-loop OFD, the linear velocity of the particles increases with the inner-loop OFD. For each set of inner-loop OFD structures, the linear velocity increases linearly as the outer-loop OFD increases. The velocity decreases for two reasons when the inner and outer ring OFD directions are opposite: first, because the positive and negative OFDs cancel each other out; second, because part of the transverse OFD transforms into revolution of dark-seeking particle.

## 3.2 Manipulation of dark- and light-seeking nanoparticles simultaneously

**Table 1. Parameters of curves to be constructed**

| Type of curve | ring | ellipse | quadrilateral |
|---|---|---|---|
| $q$ | (1,1,4,-0.5,2,2) | (1,1,2,10,6,6) | (1,1,4,10,6,6) |
| $m_1$ | -5, -5, 5, 5, -5 | -5, -5, 5, 5, -5 | -5, -5, 5, 5, -5, 5, 5, 5 |
| $m_2$ | -40, -30, 30, 40, 40 | -40, -30, 30, 40, 40 | -40, -30, 30, 40, 40, 40, 40, 40 |
| $\alpha$ | 1 | 1 | 1, 0.7, 1.2 |
| time frame(s) | 0, 12, 22, 32, 44 | 54, 64, 75, 85, 95 | 107, 117, 127, 138, 149, 160, 170, 180 |

To demonstrate the practicality of the transverse OFD designed in TC-POVB, we created three optical potential wells with varying curve shapes to manipulate both light-seeking and dark-seeking particles. These experiments are shown in Visualization 1 between 0s-160s. Additionally, we conducted further experiments between 160s-180s to adjust the OFD distribution arbitrarily, which can also be observed in Visualization 1. Table 1 presents the values of the q parameter for three curve types (circular, elliptical, and quadrilateral), along with their respective time periods and settings for m1, m2, and q values. The direction of the OFD distribution along the curve is determined by the sign of m, where positive values indicate a counterclockwise direction and negative values indicate a clockwise direction. The parameter $\alpha$ determines the uniformity of the OFD distribution along the curve. A value of $\alpha$ equals to 1 results in linear accumulation, i.e., uniform distribution. Conversely, non-zero values of $\alpha$ lead to non-uniform distribution. The first 160 seconds of Visualization 1 demonstrate the manipulation of light-seeking and dark-seeking particles moving along three different curve trajectories. The video showcases a uniform distribution of OFD along the curve, including instances of reversed OFD orientation on the inner and outer curves. From 160 to 180 seconds, the Visualization presents experiments involving particles that have a non-uniform OFD distribution.

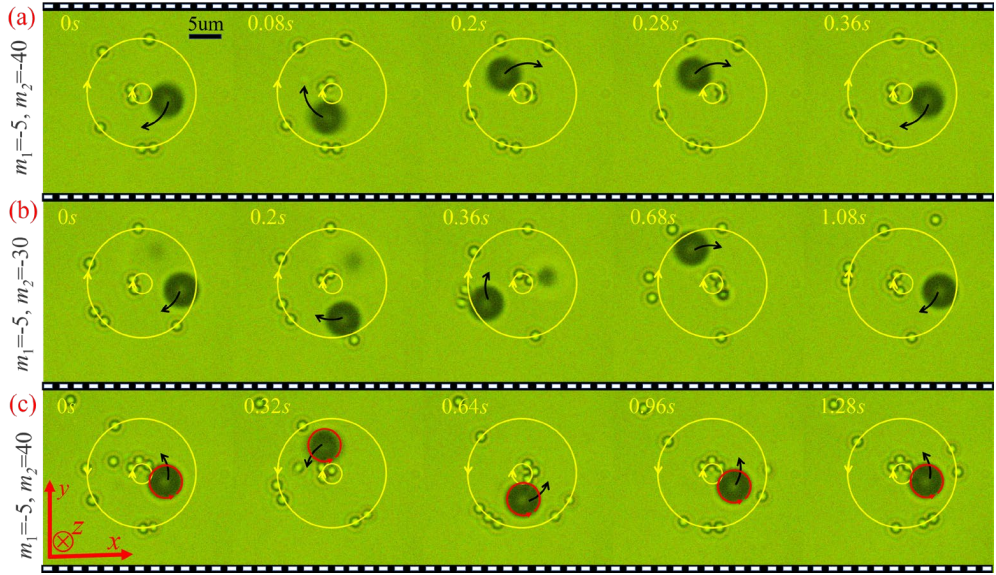

Fig. 5. Experimental results of circular TC-POVB. The first, second, and third rows show the cases of uniform OFD distribution ($\alpha = 1$) with different values of $m_1$ and $m_2$ for the inner and outer rings, respectively. The values are -5 and -40 for the first row, -5 and -30 for the second row, and -5 and 40 for the third row(see Visualization 1, 0s-54s),.

The first experimental setup is a circular TC-POVB that can propel both dark-seeking and light-seeking particles along a circular trajectory. The supplementary auxiliary lines in Figures 5, 6, and 7 offer improved clarity and comprehension. The expected motion of the dark-seeking particle is represented by black solid arrows, whereas the trajectory of the light-seeking particle is depicted by yellow solid lines with arrows indicating the direction of motion. Moreover, the direction of particle rotation is illustrated by red solid ring arrows. In Fig. 5(a), the inner ring's curve $m_1$ is -5, while the outer ring's curve $m_2$ is -40. This configuration results in clockwise OFD directions inside and outside the rings. As a consequence, both light-seeking and dark-seeking particles within the rings move in a clockwise direction, driven by the internal and external OFD. In Fig. 5(b), $m_2$ of the outer ring decreases from -40 to -30, causing a slight alteration in the intensity distribution of the inner and outer rings. The dark-seeking particles' position shifts from the center of the dark area close to the inner ring to the center of the dark area near the outer ring. Moreover, a light-seeking particle present in the aqueous solution is attracted towards the inner ring, resulting in a change from two to three light-seeking particles, as shown in Fig. 5(b). The time required to complete one cycle increases from 0.36 to 1.08 seconds since the OFD supplied by the outer ring is reduced.

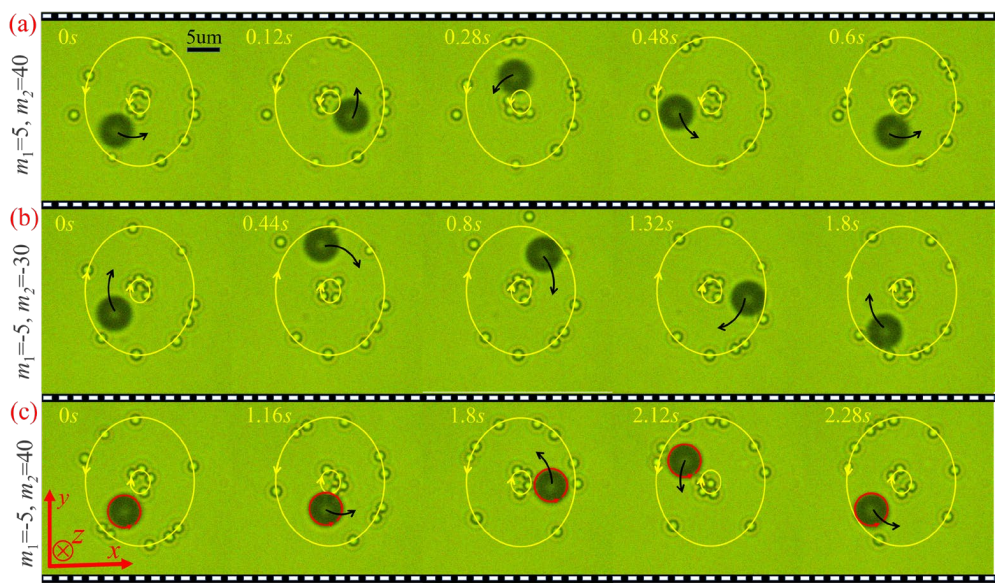

Fig. 6. Experimental results of elliptical TC-POVB. The first, second, and third rows show the cases of uniform OFD distribution ($\alpha = 1$) with different values of $m_1$ and $m_2$ for the inner and outer rings, respectively. The values are 5 and 40 for the first row, -5 and -30 for the second row, and -5 and 40 for the third row(see Visualization 1, 54s-107s).

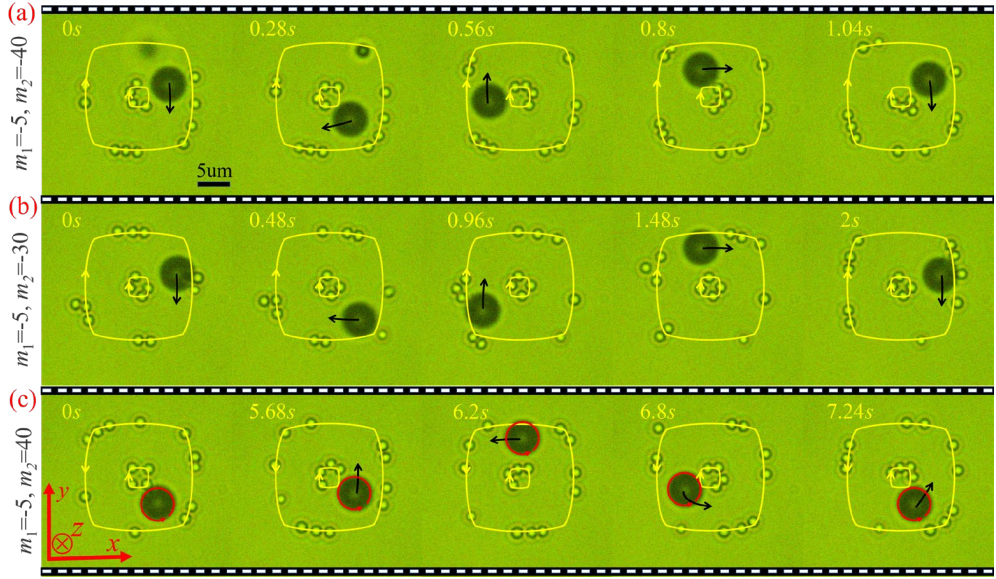

Fig. 7. Experimental results of quadrilateral TC-POVB. The first, second, and third rows show the cases of uniform OFD distribution ($\alpha=1$) with different values of m1 and m2 for the inner and outer rings, respectively. The values are -5 and -40 for the first row, -5 and -30 for the second row, and -5 and 40 for the third row(see Visualization 1, 107s-160s).

In Fig. 5(c), the OFD carried by the inner and outer rings are oriented in opposite directions. Consequently, the light-seeking particles on the inner and outer rings move clockwise and counterclockwise, respectively. The period of the dark-seeking particle further extends to 1.28 seconds. It is noteworthy that the rotation effect of the dark-seeking particle is particularly noticeable at this point due to the partial OFD transformation.

The superformula expression allows for the creation of various complex curve types. Fig. 6 and 7 depict the concurrent capture experiments of light-seeking and dark-seeking particles under elliptical and quadrilateral trajectories. In Fig. 6, the values of $m_1$ and $m_2$ in the first, second, and third rows are set to 5, 40; -5, -30; and -5, 40, respectively. In Fig. 7, they are set to -5, -40; -5, -30; and -5, 40, respectively. By comparing the inner and outer loop lateral OFD reversal cases in Fig. 4, it can be observed that the more intricate the curve is, the longer it takes for the dark-seeking particles to complete one cycle. In the third row of Figs. 5, 6, and 7, the dark-seeking particles all rotate around themselves at a certain position, particularly in the case of the quadrilateral curve trajectory in the third row of Fig. 7, where they rotate for more than 5.68 seconds before beginning to move.

The first three sets of experiments demonstrate the manipulation of light-seeking and dark-seeking particles along different trajectories. Subsequent experiments directly adjust the OFD accumulation rate by the parameter $\alpha$ while maintaining the same OFD accumulation. Figure 8(a) illustrates the linear process of OFD accumulation that drives the dark-seeking particle to move uniformly along the curve at a constant speed, as shown in Figure 8(b) when $\alpha=1$. In cases where the OFD accumulation rate is nonlinear, such as for $\alpha=0.7$ and 1.2 in Figure 8(a), corresponding to a red concave curve and a blue convex curve, respectively, the particles move fast-slow and slow-fast within the 1/4 time period intervals. As a result, the particles move from more to less and from less to more, as shown in Figure 8(c) and (d), respectively, with the black dashed line indicating the particle's position after motion in the last 1/4 time period. The moving routes are indicated by dashed lines with different colors, where black, blue, and red represent $\alpha=1$, 0.7, and 1.2, respectively.

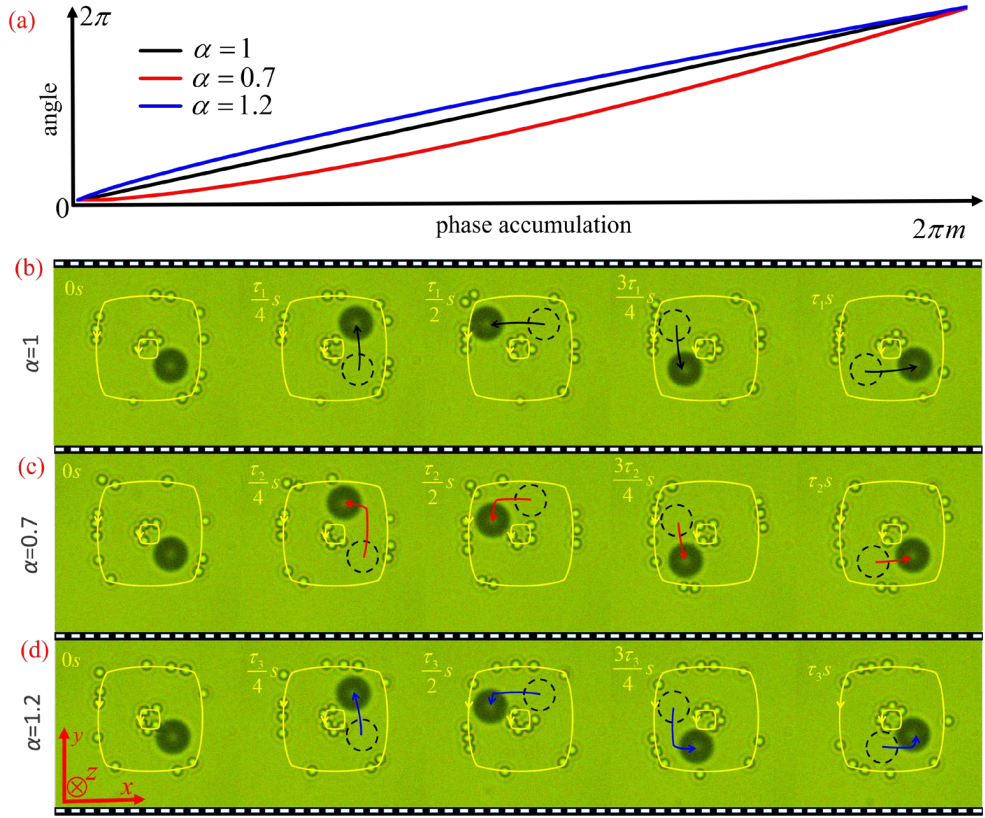

Fig. 8. Experimental results of variable speed motion. (a) the accumulation rate of OFD as a function of angle for different values of $\alpha$. (b), (c), and (d) display the motion of the particle at $\alpha$ =1, 0.7, and 1.2, respectively. The periods of dark-seeking particle motion are represented by $\tau_1$, $\tau_2$, and $\tau_3$, respectively. The parameter values m1 = 5 and m2 = 40 were utilized for the inner and outer loops in the above three sets of experiments, respectively.

## 4. Conclusion

In summary, we propose an effective scheme using OFD to manipulate both light-seeking and dark-seeking particles. Our theoretical analysis and experimental results demonstrate the efficacy of the TC-POVB beam in manipulating particle motion and how OFD affects the spin and variable-speed motion of the dark-seeking particles. These findings have broad applications and are well-suited for the development of the next generation of microfluidic tools.

**Funding**

National Science Foundation of China (12374207, 12234009, 12274215); National Key R&D Program of China (2022YFA1404800, 2018YFA0306200); Postgraduate Research & Practice Innovation Program of Jiangsu Province(KYCX23_0103).

**Disclosures**

The authors declare no conflicts of interest.